\documentclass[aps,prl,preprint]{revtex4-1}
\usepackage{amsfonts}
\usepackage{amssymb}
\usepackage{amsmath}
\usepackage{graphicx}
\usepackage{pgfplots}
\usepackage{array}

\begin{document}
\title{Spectral statistics of Toeplitz matrices }
\author{Eugene Bogomolny}
\affiliation{Université Paris-Saclay, CNRS, LPTMS, 91405 Orsay, France}
\date{\today}
\begin{abstract}
Spectral statistics of hermitian random Toeplitz matrices with independent identically distributed elements is investigated numerically. It is found  that the eigenvalue  statistics  of complex Toeplitz matrices   is surprisingly well approximated by the semi-Poisson distribution belonging to intermediate-type statistics observed in certain pseudo-integrable billiards.   The origin of intermediate behaviour   could be attributed to  the fact that Fourier transformed random Toeplitz matrices have the same slow decay outside the main diagonal as  critical random matrix ensembles. The statistical properties  of the full spectrum of real random Toeplitz matrices with i.i.d. elements are close to the Poisson distribution but each of their  constituted sub-spectra  is again well described by the semi-Poisson distribution. The  findings open new perspective in intermediate statistics. 
\end{abstract}

\maketitle 

Random matrices in physics  firstly appeared  as a tool for simplified description of complicated (and mostly unknown) quantum Hamiltonian of heavy nuclei \cite{wigner}.  Later the idea of statistical  characterisation of physical and mathematical systems with complex behaviour becomes ubiquitous  and has been successfully  applied to a huge variety of  different problems ranging from chaotic systems \cite{bohigas} and number theory \cite{keating} to quantum  gravity \cite{sachdev}, \cite{kitaev}.   Physical applications were accompanied by quick  increase  of mathematical results in random matrix theory (RMT). Besides many others it is worth to mention the exact calculations of eigenvalue distributions for invariant random matrix ensembles \cite{mehta} and the proof of their universality for wider classes of matrices \cite{tao}.  

One of the oldest class of  matrices being investigated is  Toeplitz matrices for which matrix elements $T_{mn}$ depend only on the difference of indices $m-n$
\begin{equation}
T_{m n}=a_{m-n},\qquad m,n=1,\ldots, N
\label{toeplitz}
\end{equation}
for a certain vector $a_t$ with $ t=-(N-1),\ldots, (N-1)$. 

These matrices arise naturally in many branches  of mathematics and physics, such as differential and integral equations, functional analysis, probability theory,  numerical analysis,  theory of stationary processes, signal processing, statistics,   etc. (see e.g. \cite{silbermann}, \cite{grenander} and references therein).  One of the impressive  development in this field was the asymptotic calculation of the determinants for different classes of Toeplitz matrices, starting with the classical work of Szeg\"{o}  \cite{szego} and resulting in the  proof of the Fisher-Hartwig conjecture and its generalisations (see e.g. \cite{deift} amongst others). 

Random (real symmetric) Toeplitz matrices  where $a_n$ with $n\geq 0$  are independent identically distributed  (i.i.d.)  real random variables  were introduced  in \cite{bai} where the question of their mean level density had been posed.  In \cite{bryc}, \cite{miller} it has been proved that the density of normalised eigenvalues for $N\times N$ real symmetric Toeplitz matrices for i.i.d.  $a_n$ with  zero mean value and unit variance (and  finite higher moments) converges in the limit $N\to\infty$ to a new universal distribution
\begin{equation}
\lim_{N\to\infty}\frac{1}{N}\sum_{j=1}^N\delta \Big (E-\frac{e_j}{\sqrt{N} }\Big )=\rho (E )
\end{equation} 
independent on the distribution of $a_n$. The  form of $\rho(E)$ seems to be intractable analytically.
In  \cite{virag} the behaviour of the largest eigenvalues for the above matrices has been investigated. 

Surprisingly, a very natural and important (from quantum chaos point of view) question about the spectral statistics of  random Toeplitz matrices seems not to attract attention. Only in the end of Ref.~\cite{miller} the authors briefly mentioned that in the limit $N\to\infty$ the local spacings between adjacent normalised eigenvalues for real symmetric Toeplitz matrices  should  be Poissonian (as for independent random variables).  

The main message  of this letter  is  that the spectral statistics of hermitian complex Toeplitz matrices is of different nature and it is  well approximated by the so-called semi-Poisson statistics.  The latter had been introduced in  \cite{bogomolny}  as describing   statistical properties of eigenvalues when only the nearest pairs of levels  interact as in RMT. The semi-Poisson statistics is quite special and differs from  the Poisson distribution typical for integrable models as well as from  the Wigner-Dyson statistics  appeared for chaotic systems. It   serves usually as a reference point for intermediate statistics observed in certain models. The  characteristic features of  such statistics are (i) repulsion of eigenvalues at small distances as for usual RMT and (ii) exponential decrease of the nearest-neighbour distribution at large distances as for the Poisson distribution. Such behaviour has been firstly observed in the Anderson model at the point of metal-insulator transition \cite{shklovskii} and later in certain pseudo-integrable billiards \cite{berry}, \cite{gerland}.  For example, the semi-Poisson statistics  describes well eigenvalue distribution  for a rectangular billiard with the barrier in the centre \cite{wiersig} and also appears as a particular case  of spectral statistics of interval-exchange map \cite{map} related with the Lax matrix for an integrable  Ruijsenaars-Schneider model \cite{map, giraud}. 

Two types of Toeplitz matrices  \eqref{toeplitz} are considered below:  (i) complex hermitian matrices where the real and imaginary parts of $a_n$ with $a_{-n}= a_n^*$ are i.i.d. Gaussian random variables with zero mean and  unit variance and (ii) real symmetric matrices where $a_n$  with $a_{-n}=a_n$ are real i.i.d. Gaussian random variables \cite{imaginary}. For each matrix its eigenvalues, $e_k$ and corresponding eigenfunctions, $\Psi_j(e_k)$, are calculated numerically from the diagonalisation  of Toeplitz matrices
\begin{equation}
\sum_{n=1}^N a_{m-n}\Psi_n(e_k)=e_k \Psi_m(e_k) . 
\label{toeplitz_equation}
\end{equation}
Our conclusions are  based on numerical calculations of different spectral correlation functions for these types of matrices. For the semi-Poisson (cf. \cite{bogomolny}, \cite{atas}) and the Poisson distributions they are known analytically and are presented in Table~\ref{table}. The nearest-neighbour distribution $P_n(s)$  is the probability that two eigenvalues are separated by interval $s$ and there are exactly $n$ eigenvalues inside this interval. The expressions in Table~\ref{table} correspond to the unfolded spectrum normalised to unit density, $s=\bar{d}(e_{j+1+n}-e_{j})$ where $e_j$ is an ordered set of eigenvalues and $\bar{d}$ is the mean  density. Another characteristic quantity indicated in that Table is the two-point correlation formfactor (the Fourier transform of the two-point correlation function)  which can conveniently be calculated from the unfolded spectrum by the expression
\begin{equation}
K(\tau)=\lim_{N\to\infty} \Big \langle \frac{1}{N}\left | \sum_{j=1}^N  e^{2\pi i \bar{d}  e_j\,\tau} \right |^2 \Big \rangle
\label{K_tau}
\end{equation}
where the average is taken over different realisations of the random matrix and  eigenvalues in a small window. 

To avoid the unfolding it is convenient  to calculate the probability distribution of ratio of 3 consecutive eigenvalues $r=(e_{j+2}-e_{j+1})/(e_{j+1}-e_j)$ \cite{oganesyan}. The expected formulas for this distribution are presented in the third row of Table~\ref{table}. 

\begin{figure}
\begin{center}
\includegraphics[width=.7\linewidth]{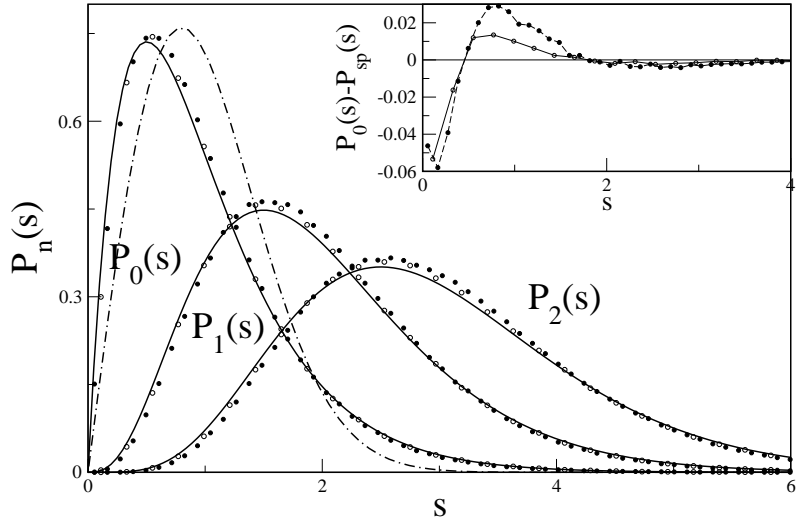}
\end{center}
\caption{ Three  nearest-neighbour distributions for complex hermitian Toeplitz matrices. Filled circles are results for $200\times 200$ matrices and $10000$ realisations. Open circles are data for $1000\times1000$ matrices and $1000$ realisations. The solid lines indicate the corresponding semi-Poisson formulas.  Dashed-dotted line is the Wigner surmise for the nearest-neighbour distribution for the usual Gaussian orthogonal ensemble: $P_{W}(s)=\tfrac{\pi}{2} s \exp(-\tfrac{\pi}{4} s^2)$. Insert: the difference between the nearest-neighbour distribution computed numerically and the semi-Poisson formula: $P_{\mathrm{sp}}(s)=4s \exp(-2s)$.}
\label{complex_ps}
\end{figure}

\begin{figure}
\begin{center}
\includegraphics[width=.7\linewidth]{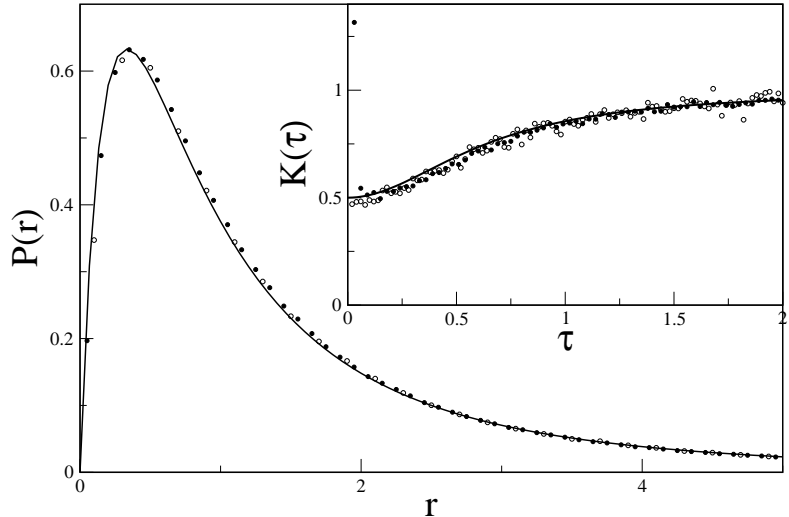}
\end{center}
\caption{ Comparison between the numerically calculated probability  distribution of the ratio of 3 consecutive eigenvalues and the semi-Poisson prediction. Insert: Comparison between the two-point formfactor and the semi-Poisson prediction. }
\label{complex_ratio}
\end{figure}

\begin{figure}
\begin{center}
\includegraphics[width=.7\linewidth]{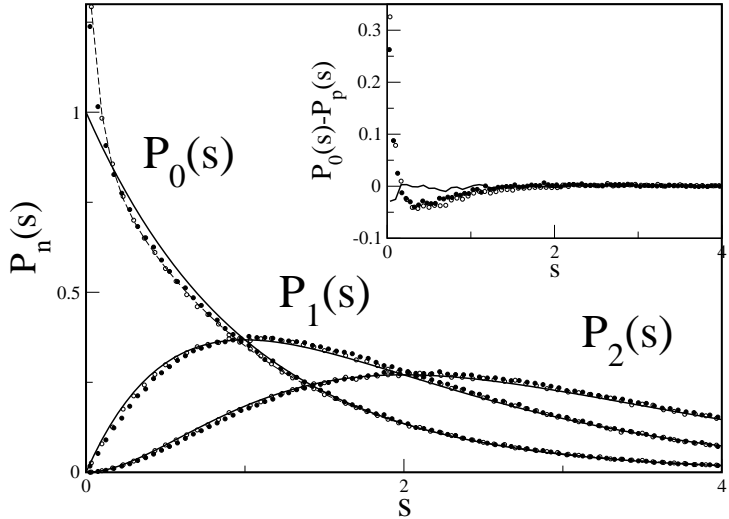}
\end{center}
\caption{The same as at Fig.~\ref{complex_ps} but for real symmetric Toeplitz matrices. Solid lines indicate the Poisson predictions to the considered quantities.  Dashed line indicates a  fit  $P_{\mathrm{fit}}(s)=0.92\, e^{-0.96 s}+0.68 \,e^{-13.7 x}$ to the nearest-neighbour distribution. Insert shows  the difference between the numerical nearest-neighbour distribution and the Poisson prediction: $P_p(s)=\exp(-s)$. Solid line in the Insert is the difference between the numerics and the above fitting formula.}
\label{real_ps}
\end{figure}

\begin{figure}
\begin{center}
\includegraphics[width=.7\linewidth]{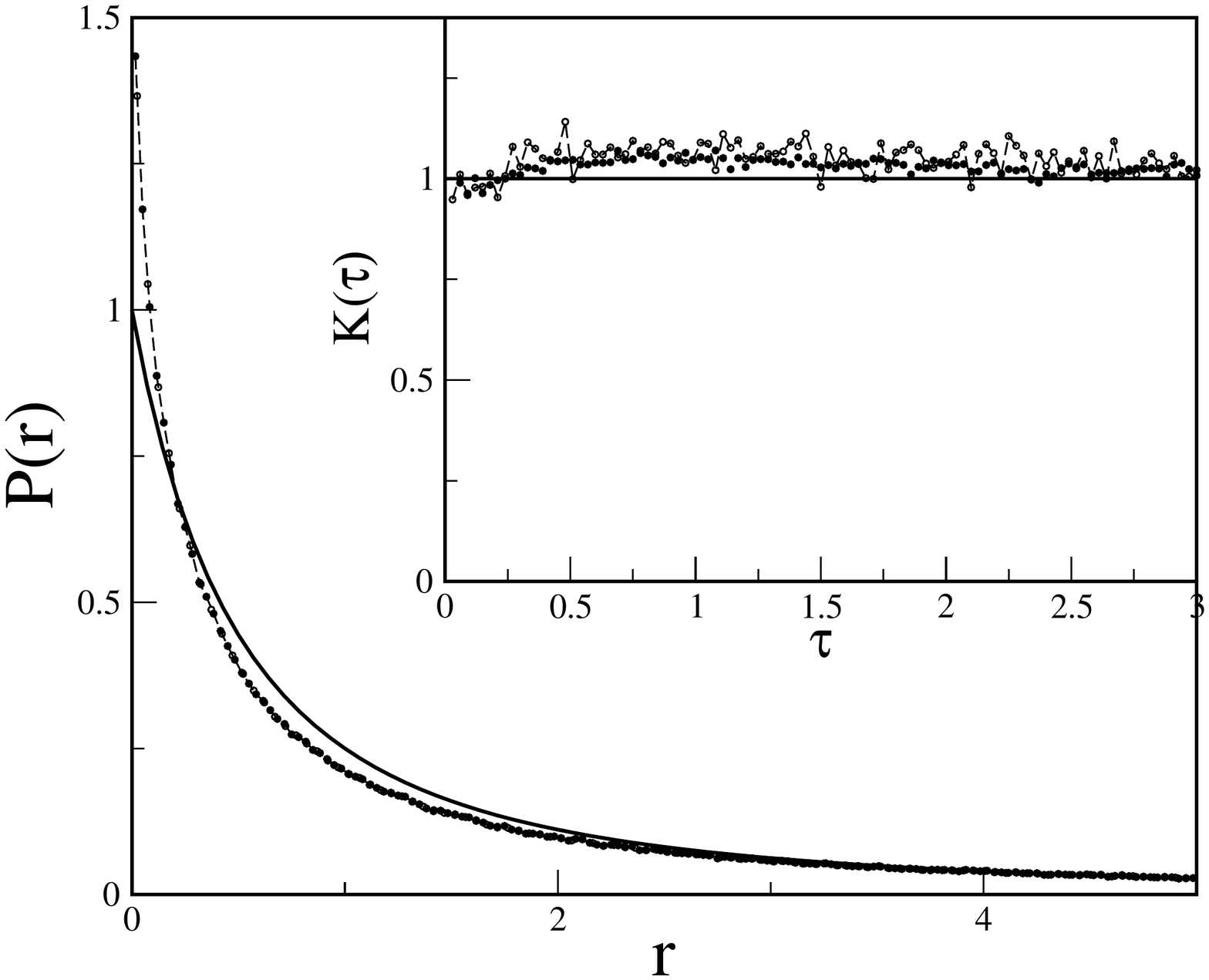}
\end{center}
\caption{The same as in Fig.~\ref{complex_ratio}  but for real symmetric Toeplitz matrices.
  Dashed line  indicates a fit $P_{\mathrm{fit}}(r)=1.51(1+5.44 r +0.90 r^2)^{-1}$ to the data. }
\label{real_ratio}
\end{figure}

The results of calculations are summarised at Figs.~\ref{complex_ps}, \ref{complex_ratio} for complex hermitian Toeplitz matrices and at Figs.~\ref{real_ps}, \ref{real_ratio} for real symmetric matrices. In each plot two sets of data are superposed. One corresponds to $200\times 200$ matrices averaged over $10000$ realisations. The second is the result of calculations for $1000\times1000$ matrices averaged over $1000$ realisations. Only minor changes were observed  with increasing the matrix dimensions and/or the number of realisations. 

At Fig.~\ref{complex_ps} the comparison between the first three nearest-neighbour distributions calculated numerically for complex hermitian Toplitz matrices and the semi-Poisson formulas is presented. In calculations only $~60\%$ of eigenvalues around the maximum of  the spectral density were taken into account and  the spectrum was unfolded by using a third degree polynomial. The agreement is quite good  and it seems that with increasing of matrix dimensions the observed distributions are  slightly closer to the semi-Poisson ones (see Insert in this figure). At  Fig.~\ref{complex_ratio} the results of calculations of the probability distribution of ratio of three nearest eigenvalues and the two-point formfactor for complex matrices are depicted. The agreement with the semi-Poisson formulas is evident. In particular, the level compressibility, $\chi=\lim_{\tau\to 0}K(\tau)$, is close to $0.5$ which is the characteristic value for the semi-Poisson distribution \cite{footnote}.  All obtained results clearly indicate that spectral correlation functions for complex Toeplitz matrices are quite well described by the semi-Poisson predictions. 

\begin{table}
\centering
\begin{tabular}{   p{ 9em }  p{9em}  p{9em} }
\hline \hline
Correlation functions & Semi-Poisson distribution & Poisson distribution \\
\hline 
$P_n(s)$ & $\dfrac{2^{2(n+1)}}{(2n+1)!}s^{2n+1}e^{-2s}$& $\dfrac{s^n}{n!}e^{-s}$\\
\hline
$K(\tau)$ & $\dfrac{2+\pi^2 \tau^2}{4+\pi^2\tau^2}$ & $1$ \\
\hline 
$P(r)$ & $\dfrac{6r}{(1+r)^4}$& $\dfrac{1}{1+r^2}$\\
\hline \hline
\end{tabular} 
\caption{Main correlation functions for the semi-Poisson and the Poisson distributions}
\label{table}
\end{table}

 At Figs.~\ref{real_ps}, \ref{real_ratio} the same  correlation functions but  for real symmetric Toeplitz matrices are plotted. In this case the results are indeed close to  the Poisson predictions  as was conjectured in \cite{miller} though  small deviations are visible. To see clearly the size of these deviations the nearest-neighbour   and the ratio distributions  were fitted by ad hoc expressions indicated in figure captions. 
 
 These  figures correspond to statistics of the full spectrum of real Toeplitz matrices \eqref{toeplitz} as it has been considered in \cite{miller}. But it is known \cite{andrew, cantoni, trench} that due to the built-in symmetry: $T_{N-n+1,N-m+1}=T_{n,m}$ eigenvectors of real symmetric Toeplitz matrices  are split  into two sub-classes, symmetric and skew-symmetric for which $\Psi_{N-m+1}^{(\pm)}=\pm \Psi_{m}^{(\pm)}$ respectively. Consequently  the spectrum of real Toeplitz matrices breaks into two sub-spectra with symmetric and skew-symmetric eigenvectors \cite{anti_unitary}. The sub-spectra can be directly  calculated from the diagonalisation  of  matrices $T^{(\pm)}$ of half-dimension \cite{andrew, cantoni, trench} with $m,n=1,\ldots, [N/2]$. For even $N$
\begin{equation}
T^{(\pm)}=  (a_{m-n}\pm a_{m+n-1} ) . 
\label{subspectrum_even}
\end{equation} 
For odd $N$ 
\begin{eqnarray}
T^{(-)}&=& (a_{m-n}- a_{m+n}) ,\label{subspectrum_odd} \\
 T^{(+)}&=& \left ( \begin{array}{cc} a_0& \sqrt{2} a_n\\ \sqrt{2} a_{m} & a_{m-n} +a_{m+n} \end{array}\right ).
 \nonumber
\end{eqnarray} 
Numerical calculations (not shown) demonstrate that the spectral correlation functions  of the both sub-spectra are also well described by the semi-Poisson statistics  and are practically indistinguishable from the ones of complex Toeplitz matrices  indicated at Figs.~\ref{complex_ps},  \ref{complex_ratio} \cite{interlacing}.   

In addition to  spectral statistics of eigenvalues of Toeplitz matrices it is of importance  to investigate statistical properties of their eigenfunctions. In usual ensembles of random matrices eigenfunctions are invariant over rotations \cite{mehta}  and in almost all coordinates they are fully extended. This is not the case for non-invariant models. Eigenfunctions of complex Toeplitz matrices (and for each sub-spectra for real matrices) in coordinate space given by \eqref{toeplitz_equation} look as fully extended. From experiences with models with intermediate statistics \cite{mirlin}, \cite{olivier}   it follows that the same eigenfunctions in the momentum representation are completely different and they may  and will have non-trivial fractal dimensions. The later are  defined  from the scaling of eigenfunction moments with matrix dimensions (see e.g. \cite{mirlin}) 
\begin{equation}
M_q=\Big \langle \sum_{n=1}^N |\Psi_n(e) |^{2q}\Big \rangle \underset{N\to \infty}{\longrightarrow} C  N^{-\tau(q)}.
\end{equation}
Here $\Psi_j(e)$ is the eigenfunction corresponded to an eigenvalue $e$ (assumed normalised) and the average is taken over a small energy window and different realisation of random matrices. The exponent, $\tau(q)$, defines the fractal dimension: 
 $D_q=\tau(q)/(q-1)$. Physically fractal dimensions determine how many important components have  eigenfunctions in a certain scale. For fully  extended eigenfunctions $D_q=1$ and for localised eigenfunctions $D_q=0$. The case when  $D_q$ depends on $q$  corresponds to multifractal eigenfunctions. 

For Toeplitz matrices \eqref{toeplitz} eigenfunctions in the momentum  representation can be calculated either by the direct Fourier transform of eigenfunctions of \eqref{toeplitz} 
\begin{equation}
\hat{\Psi}_p(e)=\frac{1}{\sqrt{N}}\sum_{j=1}^Ne^{2\pi i p j/N}\Psi_j(e)
\label{fourier}
\end{equation} 
or by the diagonalisation of  the hermitian Toeplitz matrix in the momentum  representation
\begin{eqnarray}
\hat{M}_{pr}&=&\frac{1}{N} \sum_{m=1}^N \sum_{n=1}^N a_{m-n} e^{ 2\pi i (mp-nr)/N}\nonumber\\
&=&\xi_p\delta_{pr} +(1-\delta_{pr}) \frac{2i(\eta_p-\eta_r)}{N(e^{-2\pi i (p-r)/N}-1)} 
\label{fourier_matrix}
\end{eqnarray}
where 
\begin{eqnarray}
\xi_{p}&=&a_0+2\sum_{t=1}^{N-1}\Big(1-\frac{t}{N}\Big )\mathrm{Re} \big ( a_t e^{2\pi i t p/N}\big) ,
\label{diagonal}\\
\eta_p&=&\sum_{t=1}^{N-1}\mathrm{Im}\big (a_t e^{2\pi i t p/N}\big ).
\label{off_diagonal} 
\end{eqnarray}
 Numerical determination of fractal dimensions  for the Fourier eigenfunctions \eqref{fourier}  were performed by the diagonalisation of Toeplitz  matrices of dimensions $N=2^{n}$ with $n=7,8,9,10$ averaged over $1000$ realisations.  Only eigenfunctions with eigenvalues between $-2\sqrt{N}$ and $2\sqrt{N}$  for complex matrices and between $-\sqrt{N}$ and $\sqrt{N}$ for real matrices  were taken into account.  Linear fits in the logarithmic scale work  well and permit to determine  fractal dimensions. Preliminary results  suggest that fractal dimensions are non-trivial: $D_{1/2}\approx 0.6$, $D_{3/2}\approx 0.5$, $D_2\approx 0.2$ with estimated error of order  of $0.1$ Larger-scale calculations should be performed to determine precise values of fractal dimensions for  Toeplitz matrices.   

As  Re($a_n$) and Im($a_n$) are i.i.d. Gaussian variables with zero mean and unit variance  matrix elements \eqref{diagonal} and \eqref{off_diagonal} are also Gaussian variables with zero mean and the following variances 
\begin{eqnarray}
\langle \eta_p\, \eta_r\rangle&=&N\delta_{pr}-1,\nonumber\\
\langle \xi_{p}\, \eta_{r} \rangle&=& (1-\delta_{pr}) \cot \Big ( \frac{\pi (r-p)}{N} \Big ), \label{fourier_elements}\\
\langle \xi_{p}\, \xi_{r } \rangle&=&-1+\delta_{pr}\frac{4N^2+2}{3N}+ \frac{2(1-\delta_{pr})}{N \sin^2 \big ( \frac{\pi (p-r)}{N} \big )}.\nonumber 
\end{eqnarray}
In  \cite{fyodorov} an ensemble of power law banded random matrices has been investigated. In this ensemble  each matrix element $H_{ij}$ is independent (up to the hermitian symmetry) Gaussian variable with zero mean and variance decreasing as a power from the diagonal;  $\langle H_{pr}^2 \rangle \sim (p-r)^{-2\alpha}$ $p,r\gg1$. 
It was  argued  in  \cite{fyodorov} that when $\alpha>1$ off-diagonal terms are non-essential and the spectral statistics is the same as that of diagonal elements (i.e. Poissonian). When $\alpha<1$ the spectral statistics is that of usual random matrix ensembles. Value $\alpha=1$ is special (see also \cite{levitov}) and corresponds to the so-called critical ensembles characterised  by intermediate spectral statistics similar (but in general not equal) to the semi-Poisson distribution and by non-trivial fractal dimensions calculated either numerically or in perturbation series \cite{mirlin, olivier}. 
 
A characteristic  property of matrix $\hat{M}_{pr}$ in \eqref{fourier_matrix}  is a  linear decrease of matrix elements from the main diagonal, $ \hat{M}_{pr}\sim (\eta_p-\eta_r)/(p-r)$ for $p,r\ll N$ which is precisely  the condition of   criticality (and of unusual intermediate-type spectral statistics) in random matrix  ensembles \cite{levitov}, \cite{fyodorov}. Strictly speaking matrix \eqref{fourier_matrix} does not belong to the class of matrices discussed in \cite{levitov}, \cite{fyodorov}. But physical arguments in these papers are quite general and robust and their conclusions seems to be applicable in more general setting \cite{olivier}. 

It is important to stress that the property of slow fall-off of matrix elements of Toeplitz matrices in the Fourier space is  valid only because $\eta_p$  in \eqref{fourier_elements} does not decrease for large $p\ll N $ and $N\to\infty$.  But $\eta_p$ is the Fourier transform  of $a_t$.  For any reasonably smooth function $a_t$ its high Fourier harmonics should  go to zero and matrix \eqref{fourier_matrix} would be not critical. Consequently, its spectral statistics should be the close to the statistics of the diagonal part.  

For total spectrum   of real  random Toeplitz matrices  (without the explicit splitting into two sub-spectra \eqref{subspectrum_even},  \eqref{subspectrum_odd}) such arguments do not work as their  diagonal matrix elements \eqref{diagonal}  are doubly degenerated $\xi_p=\xi_{N-p}$  (with $\eta_p=-\eta_{N-p}$) and the off-diagonal terms serve mostly to remove this  degeneracy.   Such  degenerate case was not investigated within critical banded matrix approach. Numerical results at Figs.~\ref{real_ps} and \ref{real_ratio} suggest that a certain tendency of gluing levels together  remains at least for finite $N$.

The main result of this letter  is the demonstration that spectral statistics of complex Toeplitz matrices is unusual  and in the case of i.i.d. Gaussian elements is  surprisingly well described by the semi-Poisson distribution. The root of such  intermediate type statistics is the closeness of Fourier transformed Toeplitz matrices to critical random matrix ensembles characterised by a linear decrease of matrix elements from the main diagonal. In turn, this property is related with  non-convergence of  random Toeplitz matrix symbols. Such conditions are quite robust and should be valid for different  random Toeplitz matrices but  they cannot explain the observed closeness of eigenvalue statistics of complex Toeplitz matrices with i.i.d. matrix elements  to the semi-Poisson distribution. 
In the absence of analytical results the semi-Poisson distribution should  be considered as a kind of simple Wigner-type  surmise which approximate well different spectral correlation functions of Toeplitz matrices.  

The case of real Toeplitz matrices is special. Statistics of their full spectrum  resembles  the Poisson distribution though for  relatively moderate matrix dimensions small deviations were observed. Nevertheless, spectral statistics of each of two sub-spectra for these matrices is  close to the semi-Poisson distribution.

 The intermediate  statistics till now  appeared only in certain (rare) non-generic  models. The facts that (i) Toeplitz matrices appear naturally in very different fields and (ii) their statistical properties are of  intermediate type  indicate that intermediate statistics in general and the semi-Poisson distribution  in particular are more  universal that was considered before. Further investigation of these phenomena is of considerable interest.   
\acknowledgments  
The author is greatly indebted to O. Giraud for pointing out Ref.~\cite{trench} and numerous useful discussions.



\begin{thebibliography}{99}
\bibitem{imaginary} For pure imaginary hermitian Toeplitz matrices with $a_t= i b_t$ and $b_t$ ($t=1,\ldots, N-1$) being real i.i.d. Gaussian variables the results are practically the same as for complex matrices discussed in the text.
\bibitem{anti_unitary} For complex hermitian Toeplitz matrices anti-unitary symmetry:  $T_{N-n+1,N-m+1}=T_{n,m}^*$ implies that their eigenvectors can be chosen as follows: $\Psi_{N-m+1}=\Psi_m^*$. The existence of this anti-unitary symmetry  forces the complex random Toeplitz matrices to have linear level repulsion \cite{robnik}.
\bibitem{interlacing} The two sub-spectra are not exactly interlacing and the construction of \cite{seligman} can not be applied.
\bibitem{footnote} Deviations of numerally calculated formfactor at very small $\tau$  from a $N$-independent expression  are expected as   formally $K(0)=N$ (see \eqref{K_tau}) and an increasingly large number of realisations is required to perform correctly the averaging at very small $\tau$.
\end{thebibliography}
\end{document}